# Neural Dysfunction Underlying Working Memory Processing at Different Stages of the Illness Course in Schizophrenia: A Comparative Meta-analysis

Yuhao Yao [1][#], Shufang Zhang [2][#], Boyao Wang [1], Gaofeng Zhao[3], Hong Deng[4], Ying Chen [1]*

1. Huaxi MR Research Center (HMRRC), Department of Radiology, West China Hospital of Sichuan University; Research Unit of Psychoradiology, Chinese Academy of Medical Sciences; Functional & Molecular Imaging Key Laboratory of Sichuan Province, West China Hospital of Sichuan University, Chengdu, China
2. Department of Critical Care Medicine, West China Hospital, Sichuan University/West China School of Nursing, Sichuan University, Chengdu, China
3. Department of Psychiatry, Shandong Daizhuang Hospital, Jinning, China
4. Mental Health Center, West China Hospital of Sichuan University, Chengdu, China

# Yuhao Yao and Shufang Zhang contributed equally to this work.

* Corresponding Author: Ying Chen, M.D., Ph.D., Huaxi MR Research Center (HMRRC), Department of Radiology, West China Hospital of Sichuan University, No. 37 Guo Xue Xiang, Chengdu 610041, PR China. E-mail: chenying85285@163.com

Word count: Abstract: 196; Main text: 3422

## Abstract

Schizophrenia (SCZ), as a chronic and persistent disorder, exhibits working memory deficits across various stages of the disorder, yet the neural mechanisms underlying these deficits remain elusive with inconsistent neuroimaging findings. We aimed to compare the brain functional changes of working memory in patients at different stages: clinical high risk (CHR), first-episode psychosis (FEP), and long-term SCZ, using meta-analyses of functional magnetic resonance imaging (fMRI) studies. Following a systematic literature search, fifty-six whole-brain task-based fMRI studies (15 for CHR, 16 for FEP, 25 for long-term SCZ) were included. The separate and pooled neurofunctional mechanisms among CHR, FEP and long-term SCZ were generated by Seed-based *d* Mapping toolbox. The CHR and FEP groups exhibited overlapping hypoactivation in the right inferior parietal lobule, right middle frontal gyrus, and left superior parietal lobule, indicating key lesion sites in the early phase of SCZ. Individuals with FEP showed lower activation in left inferior parietal lobule than those with long-term SCZ, reflecting a possible recovery process or more neural inefficiency. We concluded that SCZ represent as a continuum in the early stage of illness progression, while the neural bases are inversely changed with the development of illness course to long-term course.

## 1. Introduction

Schizophrenia (SCZ) is characterized as a progressive psychotic disorder, typically emerging in late adolescence or early adulthood, marked by recurrent episodes of psychosis and diminished psychosocial functioning (Haas and Sweeney, 1992; Mesholam-Gately et al., 2009). Psychophysiological deficits in working memory (WM) is one of the pivotal cognitive symptoms of SCZ, which is considered as a procedure of holding online and regulating information for short periods of time (Baddeley, 1998). Individuals with SCZ are inclined to show inferior accuracy and longer reaction time during different kinds of WM tasks covering verbal, visual, and auditory tasks, contributing to a lower level of intelligence and poorer social abilities of SCZ (Kenny et al., 1997; Piskulic et al., 2007; Seidman et al., 2012; White et al., 2010). These WM impairments present a substantial challenge for SCZ patients seeking to reintegrate into normal community life in clinical settings.

  WM dysfunctions generally observed in both the prodromal phase, known as the clinical high-risk (CHR) phase, and the onset of SCZ (Jundong et al., 2012; Lieberman et al., 2001). The CHR is characterized by functional deterioration and subtle symptoms, including attenuated psychotic phenomena and WM impairments (Zhao et al., 2022). Given the difference in the clinical performance of WM, several researches identified that CHR individuals used more time to recall fewer correct answers or order of disrupted objects relative to controls in letters and numbers tests, and displayed lower scores of Measurement and Treatment Research to Improve Cognition in Schizophrenia consensus cognitive battery in WM (Eisenacher et al., 2018; Randers et al., 2021). Weak WM was also found in psychotic patients, with repeated psychotic episodes causing progressive deterioration of WM performances (Sobizack et al., 1999; Yang et al., 2016). For example, patients with long-term SCZ performed worse, including higher error rates as well as longer response time, in WM tasks relative to patients with first-episode psychosis (FEP) (Rek-Owodziń et al., 2022; Tyburski et al., 2021; Vogel et al., 2016). However, other studies indicate that the WM impairment is not progressive after FEP, but rather stabilizes (Greenwood et al., 2008; Zanello et al., 2009) and even improves (Fu et al., 2017) in the follow-up. While there is substantial support for the neurodevelopmental model of schizophrenia in recent years (Frommann et al., 2011; Schubert and McNeil, 2007), it remains characterized by significant heterogeneity.

To further explore the brain functional alterations of WM in SCZ, increasingly more researchers have adopted neuroimaging methods, especially the functional magnetic resonance imaging (fMRI) (Zhou et al., 2022). However, noticeable heterogeneity in identified neural substrates has been observed across different stages. CHR patients showed less activation in the medial frontal cortex and parietal region than healthy controls during WM tasks (Broome et al., 2010; Curtis, 2006; Gould et al., 2003). Similarly, FEP subjects also demonstrate less activation than controls in the frontal and parietal cortices (Falkenberg et al., 2015; Niendam et al., 2014; Schneider et al., 2007; Yoon et al., 2008). For their discrepancy, researchers found that FEP patients showed poorer activation than CHR patients in these mentioned areas (Broome et al., 2010; Broome et al., 2009). Additionally, CHR patients who later converted to overt psychosis showed a distinct pattern of abnormal age-associated activation in the frontal cortex relative to non-converters (Karlsgodt et al., 2014). Therefore, during the transition to FEP, it is important to consider how changes in functional activation patterns may arise as a result of the acute onset of psychotic symptoms (Fusar-Poli et al., 2012; Giuliano et al., 2012). On the other hand, long-term SCZ patients may exhibit abnormal dorsolateral prefrontal cortex activation during WM process, with either declined or elevated activation occurring in the same region in different studies (Bor et al., 2011; Dauvermann et al., 2017; Dreher et al., 2012). Psychopathological mechanisms underlying WM deficits in FEP and long-term

SCZ patients might be similar in neuroimaging (Salgado-Pineda et al., 2018). Altogether, the findings from diverse functional neuroimaging studies can be illustrated as identifying copious WM abnormalities over the prodromal phase and onset of SCZ followed by a further more indistinct and selective alternation after the onset of psychosis (Bora and Murray, 2014).

Since previous studies focused on only one or two specific illness stages, the question of how these brain functional changes may differ over the three courses of SCZ has not been fully addressed. No systematic review or meta-analysis study compared neural dysfunction of WM across illness stages of SCZ from a neuroimaging perspective. Herein, we performed voxel-wise coordinate-based meta-analyses based on published neuroimaging studies on WM dysfunctions in patients with CHR, FEP, and long-term SCZ, and then conducted overlapping and comparative analyses between them. Our primary objective is to investigate the different or similar neural patterns in WM deficits between CHR and FEP patients, as well as those between long-term SCZ and FEP patients

## 2. Methods and Materials
### 2.1. Search Strategy and Selection Criteria

The protocol was pre-registered on the Open Scientific Framework. (Registration DOI: https://doi.org/10.17605/OSF.IO/BAJM8) We followed the Preferred Reporting Items for Systematic Reviews and Meta-analyses (PRISMA) reporting guideline in this study, which was used for systematic review and meta-analysis (Liberati et al., 2009). We searched the literature about functional neuroimaging studies on SCZ in PubMed, Embase, Web of Science, and Science Direct and selected the results for working memories studies published before May, 2023 (detailed search strategy in Online Supplementary Appendix 1). Additionally, manual searches were performed according to reference lists of previous meta-analysis studies (Picó-Pérez et al., 2022; Wu and Jiang, 2020). These are our inclusion criteria: 1) individuals with CHR, FEP, or long-term SCZ (which we defined individuals with illness duration greater than 5 years); 2) studies analyzed fMRI BOLD signal changes during tasks encoding WM at the whole-brain voxel-wise level; 3) provided the coordinates of significant clusters in a standard brain space; and 4) studies published in English in a peer-reviewed journal. And our exclusion criteria include: 1) theoretical papers, case reports, review or meta-analysis; 2) studies with individuals at genetic high risk for psychosis; 3) limited to the region of interest analysis and small volume correction; 4) no statistical comparisons on functional neural changes between individuals with schizophrenia and controls; and 5) peak coordinates and effect size were not available even after contacting the authors. All included studies were evaluated for quality and limitation (for details, see Online Supplementary Appendix 2) utilizing a 12-point Imaging Methodology Quality Assessment Checklist (Shepherd et al., 2012).

**2.2. Meta-analysis**

Voxel-wise meta-analyses of regional functional activation differences across stages were carried out using Anisotropic Effect Size Seed-based *d* Mapping (AES-SDM) software (https://www.sdmproject.com/old/). First, we used the coordinates of cluster peaks and effect sizes of significant between-group differences to set up an effect-size imputed map for each study with the anisotropic Gaussian kernels. Briefly, the t-values corresponding to each peak were transformed into Hedge's *d*, a standardized measure of effect size. Subsequently, we generated maps illustrating the distribution of *d* values and their associated variances. These individual study maps were then amalgamated to produce comprehensive meta-analytic maps. (Zhao et al., 2022). Secondly, we conducted a random-effects model to acquire the mean map for inter-group differences, and the data from each included study were combined with both positive and negative differences in the same map. Each of the three disease stages was compared with the healthy group to observe changes in WM. Voxel-wise threshold of p = .005 was applied and only neural clusters with over 20 voxels remained (Lukito et al., 2020; Radua et al., 2012). Finally, we then conducted conjunction and disconjunction analyses using the multimodal function of SDM for various disease stages to identify their significantly comparative differences or overlapping substrates (Long et al., 2022). We chose a corrected stringent threshold for testing their comparisons and overlaps at the voxel-wise $p < .0025$ and a cluster extent of 20 voxels as done in previous meta-analyses (Takeuchi et al., 2012).

### 2.3. Ancillary Analyses

We used the effect-size values extracted for each cluster at individual study to perform the meta-regression analyses for demographic variables (i.e., age and sex) in each group. Egger tests and Funnel plots were used to examine the potential publication bias (Pan et al., 2023). We also conducted the jack-knife to test the reproducibility of our findings. We repeated the main analysis for N times (N=number of datasets in the meta-analysis) and dropped one dataset at one time to test whether the results remained significant (Radua et al., 2010). And we used a statistical threshold of $p < .0005$ with a cluster extent of 20 voxels for the meta-regression analysis.

### 3. Results

### 3.1. Description of the Included Studies and Sample Characteristics

Procedure from literature search to study inclusion is presented in Figure 1. In total, 15 studies including 310 individuals with CHR (sample size-weighted mean age = 21.8 years; 66.2% male) and 277 healthy controls, 16 studies comprising 417 FEP patients (sample size-weighted mean age = 24.6 years; 71.1% male) and 450 healthy control individuals, and 25 studies including 578 individuals with long-term SCZ (sample size-weighted mean age = 35.0 years; 73.2% male) and 570 healthy controls were included in our meta-analysis. The characteristics of included articles and quality assessments were provided in Table1 (clinical measures of each stage can be found in Online Supplementary Table S1). No significant difference was found in sex ratio between three groups ($F$ = 0.306; $p$ = .737). These three groups showed significant differences in age ($F$ = 57.665; $p$ < 0.0001) in the independent-samples ANOVA test.

### 3.2. Comparisons of Illness Stages

The CHR and FEP groups showed overlapping hypoactivation corresponding to WM deficits in the right inferior parietal lobule (IPL, peak coordinates: -40, -38, 48; $z$ = -1.687; cluster size = 97), right middle frontal gyrus (MFG, peak coordinates: 48, 34, 32; $z$ = -1.661; cluster size = 90), and left superior parietal lobule (SPL, coordinates: -14, -66, 54; $z$ = -1.581; cluster size = 72, Table 2 and Figure 2). Patients with FEP displayed poorer activation than patients with long-term SCZ in left IPL (peak coordinates: -50, -46, 42; $z$ = -2.016; cluster size = 174, Table 2 and Figure 2).

### 3.3. Meta-analysis in Each Group

CHR subjects showed increased activation in the right insula (peak coordinates: 30, -8, 10; $z$ = 1.414; cluster size = 749, Table 3 and Figure 3) compared to healthy controls. In contrast, individuals with CHR displayed reduced brain function in the left superior frontal gyrus (peak coordinates: 6, 32, 50; $z$ = -1.821; cluster size = 1294), right IPL (peak coordinates: 42, -44, 48; $z$ = -1.962; cluster size = 1079) and left SPL (peak coordinates: -22, -66, 52; $z$ = -1.658; cluster size = 577, Table 3 and Figure 3) relative to healthy individuals.

Patients with FEP had higher activation in the right middle temporal gyrus (peak coordinates: 42, -66, 14; $z$ = -1.606; cluster size = 160) and lower activation in the left precuneus (peak coordinates: -14, -66, 56; $z$ = -1.815; cluster size = 174, Table 3 and Figure 3) than healthy controls.

Individuals with long-term SCZ showed increased activation in the right orbitofrontal cortex (OFC, peak coordinates: 2, 40, -22; $z$ = 1.824; cluster size

= 1048) and left MFG (peak coordinates: -30, 28, 46; $z$ = 1.663; cluster size = 224) and declined patterns in the right insula (peak coordinates: 32, 14, 2; $z$ = -2.632; cluster size = 1234, Table 3 and Figure 3) relative to healthy controls.

### 3.4. Ancillary Findings

Meta-regression analyses revealed no association between age and functional alternations in these three groups. In patients with CHR, higher proportion of male individuals was associated with lower activation in the left parahippocampal gyrus ($z$ = -2.391; $p$ < .0001, Online Supplementary Table S1). For individuals with long-term SCZ, higher proportion of male subjects was associated with greater activation in the right cerebellum ($z$ = 1.945; $p$ < .0001, Online Supplementary Table S2). Whole-brain jackknife sensitivity analyses supported the reliability and robustness of our findings (Online Supplementary Table S3). All of the brain regions with altered activation displayed low heterogeneity between studies. The Egger test indicated no potential publication bias found in the separate analysis of different illness stages (Online Supplementary Table S4), and funnel plots turned out to be symmetric across all regions.

### 4. Discussion

This comparative meta-analytic analysis ascertained the brain functional alterations underlying WM deficits in patients with SCZ across progressive stages of the illness. Both CHR and FEP individuals exhibited significant overlapping psychopathological actions, demonstrating decreased activation patterns in the frontoparietal substrates. Conversely, brain activation supporting WM of long-term SCZ subjects were more pronounced than in FEP patients, particularly in the left IPL. These shared and distinct neural mechanisms suggest that SCZ represent as a continuum in the early stage of illness progression, with the frontal and parietal regions showing similar aberrant patterns corresponding to WM deficits, while the neural bases undergo inverse changes with the development of the illness towards its long-term course.

**4.1. WM-related Neural Alterations at CHR and FEP**

CHR and FEP subjects both exhibited a decrease activation pattern in the frontoparietal regions, including the right IPL, left SPL, and right MFG. The parietal cortex is thought to be the "workspace" for sensory and perceptual processing (Andersen and Cui, 2009; Owen et al., 2005), and damage to this area leads to impaired performance on WM tasks across all stimulus categories, including spatial, object, and object/spatial conjunction tasks (Berryhill and Olson, 2008; Finke et al., 2006; Wager and Smith, 2003). Functional MRI studies reported broad bilateral activations across the posterior parietal cortex during WM tasks (Wager and Smith, 2003), reflecting attentional demands (Naghavi and Nyberg, 2005), WM maintenance (Song and Jiang, 2006; Xu and Chun, 2006), and information accrual (Xu, 2007). Posterior parietal cortex, particularly the IPL and SPL, integrates information from visual, auditory and sensory modalities, and is also involved in the evaluation of information and planning of response (Pearlson et al., 1996; Torrey, 2007). The MFG, acting as a convergence site for the dorsal and ventral attention networks (Japee et al., 2015), plays a key role in maintaining information and supporting executive function during WM (Bokde et al., 2010). Our findings align with these roles, as the three identified brain regions belong to the frontoparietal network, which is a crucial component of the WM network (Marek and Dosenbach, 2018) (Dosenbach et al., 2008; Dosenbach et al., 2007; Marek et al., 2015). Reduced brain activity and connectivity within and between regions of the frontoparietal network have been demonstrated in SCZ across various WM tasks (Anticevic et al., 2013; Cole et al., 2011; Sheffield et al., 2015). This common hypoactivation and reduced integration can be summarized as a dysfunction of the frontoparietal network, which is a shared characteristic between CHR and FEP. In conclusion, our findings imply shared triggers for dysfunctional WM in CHR and FEP, providing a novel therapeutic target for improving functional impairment in the early stages of SCZ.

Our findings of hyperactivation in the insula of individuals with CHR and in temporal gyrus of patients with FEP are in line with the available functional research in CHR or FEP reporting similar brain regions. The insula plays as a component of the limbic integration cortex (Augustine, 1996; Takahashi et al., 2009), engaging in the maintenance process of WM (Schneider et al., 2007). Individuals with CHR are noted to have a smaller insular cortex compared to health controls, yet increasing activation is observed during WM tasks, suggesting a compensatory response to anatomical abnormalities. The middle temporal gyrus, although not directly related to WM, was found to deactivate with increasing WM load in healthy individuals. For patients with FEP, the failure to deactivate the temporal regions during WM tasks that engage frontal cortex reflects a disruption of frontotemporal connectivity (Crossley et al., 2009). This hyperactivation in the middle temporal gyrus may be a reversed coupling with the frontal region, indicating an alteration in microcircuitry stemming from abnormal frontal activity (Friston and Frith, 1995).

### 4.2. WM-related Neural Alterations Associated with Long-term SCZ

In the direct comparison between FEP and long-term SCZ, Individuals with FEP showed hypoactivation in left IPL relative to long-term SCZ patients in our study. The IPL is thought to be the core domain of the executive control network, responsible for representing retrieved information into corresponding sensory areas (Kahn et al., 2004; Wheeler et al., 2000). Dysfunction in IPL has been found in patients with SCZ, with differentially expressed diacylglycerol kinase causing WM impairment (Liu et al., 2023). As WM involves distinct functional stages, such as encoding, maintenance, retrieval (Cairo et al., 2004; Manoach, 2003; Woodward et al., 2006), abnormal neural activity in the IPL have been associated with deteriorated encoding and maintenance abilities (Choi et al., 2012; Meda et al., 2009). Our finding indicates a potential recovery of WM-related function in the long-term development of SCZ. One possible explanation for this improvement is the promoting effect of medication on functional recovery in SCZ. Antipsychotic medication has been shown to ameliorate cognitive impairment (Davidson et al., 2009; Désaméricq et al., 2014; Schooler et al., 2005) and may influence the Blood-Oxygen-Level-Dependent signal in specific regions during WM process (Peitl et al., 2021; Röder et al., 2013). A resting-state fMRI study indicated increased low-frequency fluctuation in the IPL among SCZ patients after antipsychotic treatment (Wang et al., 2021), implying a specific activation effect of atypical antipsychotics. On the other hand, over-recruitment of IPL might represent that long-term SCZ causes more neural inefficiency than FEP. Several studies suggested that task performances in WM deteriorate over the course of the illness, and WM abnormalities are more pronounced in the long-term SCZ patients than in FEP patients (McCleery et al., 2014; Wu et al., 2016). Given these results, long-term SCZ devotes greater cortical resources in the IPL under increasing demand but accomplishes worse levels of accuracy and response time to those of FEP during WM process, implying degradation of neural inefficiency in the process of disorder chronicity.

Compared with healthy controls, individuals with long-term SCZ showed hyperactivation in the right OFC and declined patterns in the right insula. Orbitofrontal lesions, which occur in SCZ (Castro-Fornieles et al., 2018; James et al., 2011), are associated with deficits in executive control functions underlying the joint maintenance, manipulation, and monitoring of information in working memory, particularly at high levels of WM load (Barbey et al., 2011). The greater activation in OFC might represent compensatory mechanisms to address decreased efficiency in achieving a similar level of WM processing as healthy controls. Our finding of hypoactivation in the insula can plausibly account for the deficits in WM, particularly in maintenance process (Schneider et al., 2007). Additionally, both OFC and insula belong to the paralimbic system (Mesulam, 2000; Mesulam, 1998), which might imply the dysconnectivity of brain regions in limbic system as the disorder progresses, shedding light on underlying link between WM and limbic system.

### 4.3. Limitations

Our study has shortcomings that need to be considered. First, while our meta-analysis provides the opportunity to aggregate cross-sectional studies from the existing literature to assess the potential regional pattern of progressive functional changes, further longitudinal studies may be required to validate our findings. Second, we were unable to investigate the medication effects comprehensively due to the lack of precise dosage information. It is important to acknowledge that antipsychotic medications may potentially play a role in the observed group differences especially within the long-term SCZ group. Third, considerable heterogeneity exists among individuals meeting CHR criteria, with varying outcomes – some progressing to psychosis, while others do not, and some developing conditions other than SCZ. Identifying distinct heterogeneity in CHR individuals, particularly in relation to brain features and clinical outcomes, may enhance our understanding of the similarities and disparities among CHR subgroups in comparison to individuals with FEP. Finally, inconsistent tasks were another prominent source of heterogeneity. Although the majority of included studies employed the n-back task, some utilized verbal forms or the Sternberg task to assess WM.

### 4.4. Conclusion

This comparative meta-analysis found that neural dysfunctions underlying WM processing were highly prevalent throughout the course of SCZ. Different stages of SCZ exhibited both shared and distinct deficits in brain function. The overlapping brain regions in CHR and FEP suggest important triggers for incipient WM dysfunction, extending potential therapeutic target for improving functional impairment in early SCZ. Additionally, we observed increased activation in patients with long-term SCZ in left IPL compared with FEP and may reflect a recovery process or more neural inefficiency. Altogether, these findings support understanding of the WM performance of SCZ manifested at specific stages of illness and expand characterization of increasing illness-related alterations in brain function over the illness course.

## Acknowledgments

We acknowledge and appreciate the efforts of all the authors of the included studies who responded to our requests for further information not included in published manuscripts.

## Data Availability Statement

All the data that support the findings of the present study are available from the corresponding author through request.

## Financial Disclosure

This work was supported by the National Natural Science Foundation (81801358), the Key research and development project of science and technology, department of Sichuan Province (2022YFS0179) and the Key research and development project of science and technology, department of Chengdu (2022YF0501507SN). Medical and Health Science and Technology Development Plan of Shandong Province (202103090715).

## Declaration of Interest Statement

The authors have declared that no conflict of interest exists.

## Author Contributions

YY and YC designed the study. YY and SZ collected the data from previous studies. YY, SZ, and BY performed the data analysis and wrote the paper. ZG and DH helped with data processing. BY, YC revised the paper. All authors contributed to the results' interpretation and discussion and approved the final manuscript.

Comparative Meta-analysis Across Illness Stages of SCZx

**Table 1. Sample Characteristics and Summary Findings of Included Studies**

| Study | Patient Group | | | Control group | | | Scan./FWHM (mm) | Slice thick. (mm) | Cor | Quality score | Task | Main finding |
|---|---|---|---|---|---|---|---|---|---|---|---|---|
| | N. | Age | % Male | N. | Age | % Male | | | | | | |
| **CHR** | | | | | | | | | | | | |
| Broome 2009.[55] | 17 | 24.2±4.1 | 70.6% | 15 | 25.4±4.9 | 73.3% | 1.5T/ 7.2 | 3 | N | 10 | N-back | Controls > CHR：PCUN, IPL, calcarine, INS, IFG, SPL, anterior cingulate, SFG |
| Broome 2010.[9] | 17 | 24.2±4.1 | 70.6% | 15 | 25.4±4.9 | 73.3% | 1.5T/ 7.2 | 3 | N | 10 | Spatial working memory task | Controls > CHR: MFG, SFG, PCUN, Cereb， |
| Choi 2012.[29] | 21 | 21.6±4.1 | 57.1% | 16 | 21.4±2.3 | 56.3% | 1.5T/ 8 | 5 | N | 11 | Spatial working memory task | Controls > CHR: IPL, posG, DLPFC, Putamen, ITG, SPL, IPL, VLPFC CHR > Controls: STG, MTG, INS, ACC, DLPFC, MFG, preG, posG, ITG, PHG, PCUN |
| Crossley 2009.[84] | 16 | NA | NA | 13 | NA | NA | 1.5T/ 6 | 3 | Y | 11 | N-back | NS |
| Falkenberg 2015.[9] | 17 | 24.3±4.2 | 70.6% | 15 | 25.6±4.8 | 66.7% | 1.5T/ 10 | 8 | Y | 11 | N-back | CHR > Controls：LING, PCUN |
| Falkenberg 2016.[2] | 34 | 22.0±NA | 70.6% | 20 | 25.5±NA | 60.0% | 1.5T/ NA | 7 | Y | 11.5 | Delayed matching to sample task | CHR > Controls：IFG Controls > CHR：CUN, PCG |
| Fusar-Poli 2010.[107] | 20 | NA | NA | 14 | NA | NA | 1.5T/ 6 | 3 | Y | 10 | N-Back | Controls > CHR: MFG, MFG, SPL |

| Study | N | Age | % Male | N | Age | % Male | Field/Slices | Threshold | Y/N | Smoothing | Task | Findings |
|---|---|---|---|---|---|---|---|---|---|---|---|---|
| Fusar-Poli 2010.[26] | 15 | NA | NA | 15 | NA | NA | 1.5T/6 | 3 | Y | 9.5 | Visuospatial working memory task | Controls > CHR：PCUN, SPL, MTG |
| Fusar-Poli 2011.[42] | 15 | NA | NA | 15 | NA | NA | 1.5T/6 | 3 | Y | 9.5 | N-Back | Controls > CHR: SMG, MFG, IPL |
| Karlsgodt 2014.[9] | 20 | 16.9±2.1 | 85.0% | 19 | 17.8±2.1 | 52.7% | 3T/5 | 3 | Y | 11 | Sternberg task | CHR > Controls: MFG, frontal eye fields, CG, putamen, frontal pole, INS, OFC, angular gyrus, Tempero-occipital region, SMA |
| Niendam 2014.[25] | 25 | 16.9±3.9 | 56.0% | 35 | 17.6±3.2 | 54.0% | 1.5T/8 | 4 | Y | 11.5 | AX-CPT | Controls > CHR：preG, MFG, MCG, IPL |
| Pauly 2010.[11] | 12 | 24.2±4.6 | 83.3% | 12 | 24.5±4.7 | 100.0% | 1.5T/8 | 3 | Y | 11 | N-Back | Controls > CHR: SPL, PCUN, postcentral lobe; CHR > Controls: preG, INS |
| Schmidt 2014.[20] | 27 | 25.0±5.0 | 74.1% | 19 | 36.5±4.0 | 52.6% | 3T/8 | 3 | Y | 10.5 | N-back | Controls > CHR: MFG, SPL |
| Smieskova 2012.[33] | 17 | 25.2±6.3 | 76.5%% | 20 | 26.5±4.0 | 50.0% | 3T/8 | 3 | Y | 10 | N-back | Controls > CHR: SPL, PCUN, SFG, IPL, |
| Thermenos 2016.[10] | 37 | 19.6±4.0 | 41.0% | 34 | 20.0±4.1 | 53.0% | 3T/8 | 3 | Y | 10 | N-Back | CHR > Controls: PHP |
| **FEP** | | | | | | | | | | | | |
| B`egue 2022.[1] | 23 | 23.6±1.4 | 87.0% | 27 | 33.1±1.7 | 63.0% | 3T/6 | 3 | Y | 10.5 | N-back | FEP > Controls：Cereb, MFG |
| Borgan 2021.[7] | 31 | 26.6±4.7 | 83.9% | 35 | 27.1±5.3 | 74.3% | NA | NA | Y | 11 | Sternberg task | Controls > FEP：HIP, PCG, Parietal operculum; FEP > Controls: Angular gyrus, MOG, |

| Study | n | Age | % Male | n | Age | % Male | Field/Slice | Threshold | Med | Duration | Task | Findings |
|---|---|---|---|---|---|---|---|---|---|---|---|---|
| | | | | | | | | | | | | SPL, Occipital FUS |
| Broome 2009.[55] | 10 | 25.5±5.9 | 70.0% | 15 | 25.4±4.9 | 73.3% | 1.5T/7.2 | 3 | N | 10 | N-back | Controls > FEP: IPL, PCUN, calcarine, INS, IFG, SFG, SPL, ACC, |
| Broome 2010.[9] | 10 | 25.5±5.9 | 70.0% | 15 | 25.4±4.9 | 73.3% | 1.5T/7.2 | 3 | N | 10 | Spatial working memory task | Controls > FEP: MFG, SFG, PCUN, Cereb |
| Lesh 2013.[64] | 43 | NA | 79.0% | 54 | NA | 65.0% | 1.5T/8 | 4 | Y | 9 | AX-CPT | Controls > FEP: IPC, DLPFC |
| Crossley 2009.[84] | 10 | NA | NA | 13 | NA | NA | 1.5T/6 | 3 | Y | 9 | N-back | NS |
| Lesh 2015.[66] | 45 | 20.3±13.1 | 82.0% | 37 | 19.7±2.6 | 73.0% | 1.5T/10 | 4 | Y | 10 | AX-CPT | Control > FEP: DLPFC, IPC |
| Nejad 2011.[14] | 23 | 26.2±5.0 | 78.3% | 35 | 26.8±5.8 | 68.6% | 3T/8 | 3 | Y | 11 | N-back | FEP > Controls: INS, STG, IPG. Controls > FEP: INS, rolandic operculum |
| Niendam 2014.[25] | 35 | 18.3±2.6 | 74.0% | 35 | 17.6±3.2 | 54.0% | 1.5T/8 | 4 | Y | 11.5 | AX-CPT | Controls > FEP: IPL |
| Scheuerecker 2008.[13] | 23 | 31.6±11.1 | 82.6% | 23 | 32.6±9.9 | 82.6% | 1.5T/8 | 1.5 | Y | 10.5 | N-back | Controls > FEP: CAU, INS; FEP > Controls: FUS, Rolandic operculum |
| Schneider 2007.[55] | 48 | 31.0±9.9 | 54.2% | 57 | 30.9±8.3 | 54.4% | 1.5T/10 | 3 | Y | 11 | N-back | Controls > FEP: PCUN, posG, FEP > Controls: VLPFC, INS, MOG |
| Tan 2005.[50] | 11 | 25.0±5.5 | 45.5% | 11 | 25.9±6.4 | 45.5% | 3T/8 | 3 | N | 10.5 | Sternberg task | Controls > FEP: MFG, ACC, INS, THAL; FEP > Controls: IFG, MFC, IPL, preG, INS |
| van Veelen 2010.[25] | 30 | 24.7±4.2 | 100.0% | 36 | 24.3±4.6 | 100.0% | 1.5T/8 | 4 | Y | 10 | Sternberg tasks | FEP > Controls: DLPFC |

| Yoon 2008.[116] | 25 | 19.6±3.8 | 68.0% | 24 | 21.6±4.24 | 54.0% | 1.5T/8 | 4 | Y | 11 | AX-CPT | Controls > FEP: MFG, IFG |
| Zhang 2013.[18] | 33 | 23.8±5.8 | 48.5% | 15 | 24.1±6.4 | 40.0% | 3T/8 | 3 | Y | 10.5 | N-back | FEP > Controls: MPFC, PCC |
| Zhou 2014.[7] | 17 | 23.7±6.9 | 58.8% | 18 | 24.9±7.3 | 50.0% | 3T/6 | 4 | Y | 10 | N-back | Controls > FEP: PCUN, ITG, IFG, MFG |

| Study | Patient Group | | | | Control group | | | Scan/FWHM (mm) | Slice thick. (mm) | Cor | Quality score | Task | Main finding |
|---|---|---|---|---|---|---|---|---|---|---|---|---|---|
| | N. | Age | %Male | Time | N. | Age | %Male | | | | | | |
| **Long-term** | | | | | | | | | | | | | |
| Avsar 2011.[7] | 10 | 31.4±10.3 | 100.0% | 13.3±10.8 | 8 | 35.4±15.4 | 50.0% | 1.5T/8 | 5 | Y | 10.5 | Delayed matching to sample task | Controls > SCZ: MTG, MOG, posG, PC, preG, MFG, Globus pallidus, Culmen, THAL, INS, MCG, STG, PCUN, FUS, ITG, Caudate; SCZ > Controls: MFG, IPL, preG, IPL, Dentate, Cuneus |
| Barch 2002.[55] | 38 | 36.3±10.3 | 63.0% | 13.0±1.5 | 48 | 36.5±11.2 | 46.0% | 1.5T/8 | 8 | N | 10 | N-back | Controls > SCZ: HIP, brain stem, basal ganglia, THAL, Medial superior prefrontal cortex, somatosensory cortex |
| Becerril 2011.[16] | 38 | 36.7±9.1 | 66.0% | 17.4±11.2 | 32 | 36.2±10.9 | 66.0% | 3T/9 | 3 | N | 9.5 | N-back | SCZ > Controls: Cereb, posG |
| Bor 2011.[16] | 22 | 28.4±7.2 | 77.3% | 11.9±2.4 | 15 | 30.3±7.3 | 73.3% | 1.5T/8 | 4 | Y | 11 | Spatial and verbal working memory task | SCZ > Controls: THAL, MFG |
| Cairo 2006.[22] | 15 | 32.6±11.0 | 66.7% | 10.1±NA | 15 | 32.4±10.1 | 66.7% | 1.5T/8 | 5 | Y | 10 | Sternberg task | NS |
| Dauvermann 2017.[7] | 15 | 37.1±10.0 | 86.7% | 21.5±6.1 | 18 | 37.1±10.0 | 72.2% | 3T/8 | 4 | Y | 10 | N-back | Controls > SCZ: DLPFC, IPS, ACC, VTA |
| Deserno 2012.[90] | 41 | 34.1±10.4 | 75.6% | 5.8±6.8 | 42 | 35.4±12.3 | 54.8% | 1.5T/8 | 5.5 | Y | 10.5 | N-back | Controls > SCZ: OFC |
| Diaz 2011.[13] | 11 | 32.6±12.7 | 90.9% | 13.7±14.5 | 17 | 24.0±3.9 | 41.2% | 3T/5 | 3.8 | Y | 10 | Verbal working memory task | Controls > SCZ: posterior cingulate, parietal cortex, OFC |

| Study | N | Age | % Male | Illness duration | N | Age controls | % Male controls | Scanner/slices | Smoothing | Med | Threshold | Task | Findings |
|---|---|---|---|---|---|---|---|---|---|---|---|---|---|
| Dreher 2012.[14] | 17 | 30.7±6.3 | 58.8% | 9.0±7.0 | 19 | 27.5±5.0 | 78.0% | NA/6.5 | NA | Y | 9.5 | N-back | Controls > SCZ: MFG, fronto-polar cortex, Premotor cortex, Pre-SMA, THAL, Cereb<br>SCZ > Controls: anterior mPFC, PCC, STG, PHG |
| Ettinger 2011.[16] | 45 | 37.3±8.2 | 77.8% | 13.5±2.3 | 19 | 33.3±9.2 | 63.2% | 1.5T/8 | 7 | Y | 10 | N-back | SCZ > Controls: occipital cortex, MFG, IFG, |
| Hamilton 2009.[31] | 21 | 32.4±9.5 | 70.0% | 8.8±7.8 | 38 | 32.5±11.7 | 60.5% | 3T/8 | 2.19 | Y | 11 | Delayed matching to sample task | Controls > SCZ: INS, IFG, STG, VLPFC |
| Hashimoto 2014.[7] | 17 | 31.1±6.2 | 53.0% | 7.1±6.1 | 17 | 27.7±4.3 | 76.0% | 1.5T/8 | 4 | Y | 10 | N-back | Controls > SCZ: SFG, SMG, Cuneus, PCUN, Middle cingulate |
| Honey 2002.[24] | 20 | 34.6±6.6 | 100.0% | 11.8±7.3 | 20 | 39.0±13.6 | 100.0% | 1.5T/7 | 7 | N | 9 | Verbal working memory task | NS |
| Honey 2003.[15] | 30 | 36.9±9.2 | 90.0% | 10.2±10.0 | 27 | 35.1±9.9 | 77.8% | 1.5T/NA | 7 | N | 9 | Verbal working memory task | Controls > SCZ: dorsolateral prefrontal cortex, preG |
| Hugdahl 2004.[51] | 12 | 32.4±8.0 | 50.0% | 8.7±8.2 | 12 | 31.0±5.5 | 41.7% | 1.5T/8 | 3 | Y | 10 | Mental arithmetic task | SCZ > Controls: IPL, cingulate gyrus<br>Controls > SCZ: IFG, supramarginal, lingual gyri |
| Kang 2011.[30] | 12 | 45.4±11.3 | 83.0% | 22.0±10.2 | 11 | 44.0±7.9 | 73.0% | 3T/7 | 3.5 | N | 9,.5 | Spatial working memory task | Controls > SCZ: DLPFC, IFC, INS, SMA, primary visual cortex, Cereb, THAL, putamen, HIP |
| Kim 2010.[7] | 12 | 40.2±10.2 | 58.3% | 14.1±9.9 | 13 | 40.4±9.3 | 61.5% | 3T/4 | 1 | Y | 10 | Delayed matching to sample task | SCZ > Controls: MFG, SFG |
| Manoach 2005.[12] | 16 | 42.0±11.0 | 87.5% | 18.0±10.0 | 12 | 35.0±10.0 | 66.7% | 3T/8 | 5 | N | 10 | Spatial and shape working memory task | Controls > SCZ: claustrum, INS, Posterior cingulate, HIP, Paracentral lobule, CG, Caudate, MFG, posG<br>SCZ > Controls: Cingulate g., STG, IFG, MFG |
| Matsuo 2013.[9] | 46 | 31.5±7.0 | 50.0% | 7.7±5.2 | 46 | 31.9±9.8 | 50.0% | 3T/8 | 1 | Y | 10 | Verbal working memory task | Controls > SCZ: THAL, Vermis, pons, IFG<br>SCZ > Controls: medial prefrontal cortex, angular gyrus, MTG, MCC, PCC, MFG, SMG, posG, ITG, FUS, MOG |
| Perlstein 2001.[197] | 17 | 36.5±7.5 | 64.7% | 13.9±8.4 | 16 | 36.5±6.9 | 62.5% | 1.5T/8 | 3.75 | N | 9.5 | N-back | SCZ > Controls: ACG, MFG |

| Study | N | Age | % Male | Duration | N (HC) | Age (HC) | % Male (HC) | Scanner/FWHM | Threshold | Medicated | Education | Task | Findings |
|---|---|---|---|---|---|---|---|---|---|---|---|---|---|
| Pomarol-Clotet 2008.[127] | 32 | 41.6±8.8 | 65.6% | 21.8±9.1 | 32 | 41.0±11.0 | 65.6% | 1.5T/NA | 7 | Y | 10.5 | N-back | Controls > SCZ: Cereb, basal ganglia, THAL, middle lateral PFC<br>SCZ > Controls: ACG, anterior temporal regions |
| Schlagenhauf 2008.[24] | 10 | 34.6±12.9 | 80% | 7.6±9.4 | 10 | 33.8±12.5 | 80.0% | 1.5T/8 | 1 | Y | 10 | N-back | Controls > SCZ: DLPFC, |
| Schlosser 2008.[34] | 41 | 30.2±9.5 | 68.3% | 5.0±6.8 | 41 | 29.2±8.9 | 65.9% | 1.5T/8 | 1 | Y | 10 | Sternberg task | Controls > SCZ: IFG, MFG, SFG, ACG, THAL<br>SCZ > Controls: PCUN, IPL, IFG, MFG |
| Stäblein 2019. | 25 | 36.8±9.4 | 68.0% | 11.4±8.0 | 25 | 34.9±10.5 | 48.0% | 3T/NA | 3 | Y | 11 | Visual working memory task | Controls > SCZ: PrCG, MeFG, MFG, IFG, PoCG, IPL, PCUN, MTG, STG, FG, MOG, CUN, LG, CG, PHG, HIP, THAL, CAU, claustrum, clumen<br>SCZ > Controls: ACC, |
| Walter 2007.[35] | 15 | 33.1±6.5 | 66.7% | 6.1±5.3 | 17 | 30.9±8.8 | 52.9% | 1.5T/9 | 4 | N | 9.5 | Sternberg task | Controls > SCZ: putamen, CAU, SFG, SPL, IPL, Cereb<br>SCZ > Controls: STG |

*Abbreviation:* CHR, Clinical High Risk; FEP, first episode non-affective psychosis; FWHM, Full Width at Half Maximum; PCUN, precuneus; IPL, inferior parietal lobule; OFC, orbitofrontal cortex; STG, superior temporal gyrus; IFG, inferior frontal gyrus; HIP, hippocampus; Cereb, cerebellum; SFG, superior frontal gyrus; MTG, middle temporal gyrus; PCC, posterior cingulate cortex; MOG, middle occipital gyrus; INS, insula; ACC, anterior cingulate cortex; MFG, middle frontal gyrus; NS, no significance; THAL, thalamus; SPL, superior parietal lobule; posG, postcentral gyrus; MCC, median cingulate gyrus; SMG, supramarginal gyrus; preG, precentral gyrus; SMA, supplementary motor area; FUS, fusiform gyrus; MeFG, medial frontal gyrus; FG, fusiform gyrus; PHG, Para hippocampal gyrus; CAU, caudate; ACG, anterior cingulate gyrus; ITG, inferior temporal gyrus; PCG, posterior cingulate gyrus; PHP, parahippocampus; IPS, intra-parietal sulcus; VTA, ventral tegmental area; CUN, cuneus; LG, lingual gyrus; CG, Cingulate gyrus.

**Table 2. Overlapping and Comparative Results of Different Illness Stages**

| Contrast /Brain region | Brodmann area | Coordinate (x, y, z) | SDM-Z | p value | Cluster size | Cluster breakdown (Number of voxels) |
|---|---|---|---|---|---|---|
| *Overlapping: CHR & FEP* | | | | | | |
| R inferior parietal lobule | 2 | 40, -38, 48 | -1.687 | .0002 | 97 | R inferior parietal lobule (67); R supramarginal gyrus (19); R postcentral gyrus (13) |
| R middle frontal gyrus | 45 | 48, 34, 32 | -1.661 | .0002 | 90 | R middle frontal gyrus (48); R inferior frontal gyrus, triangular part (42) |
| L superior parietal lobule | 7 | -14, -66, 54 | -1.581 | .0003 | 72 | L superior parietal lobule (40); L precuneus (24) |
| *Comparative: FEP & Long-term SCZ* | | | | | | |
| L inferior parietal lobule | 40 | -50, -46, 42 | -2.016 | <.0001 | 174 | L inferior parietal lobule (166) |

*Note:* Suprathreshold clusters are identified at p < .0005 and cluster size > 50 voxels for the overlapping and comparative analysis. The multimodal tool of the SDM is used to uncover the overlapping neural substrates across stages, and the findings only represent an overlap of the significant findings of the two separate meta-analyses within priori regions (SDM adjusts the raw union of probabilities to the desired threshold to avoid the likelihood that the false positive rate is higher than a preferred degree in the worst-case scenario). Peak coordinates were reported in the standard template of Montreal Neurological Institute (MNI), and the number of cluster breakdowns (> 10 voxels) was calculated by adding subclusters reported by SDM software. Abbreviations: L = left; R = right; CHR = clinical high-risk state for psychosis; FEP = first-episode psychosis; SCZ = schizophrenia.

**Table 3. Results of the Separate Meta-analysis of Different Illness Stages**

| Contrast /Brain region | Brodmann area | Coordinate (x, y, z) | SDM-Z | p value | Cluster size | Cluster breakdown (Number of voxels) |
|---|---|---|---|---|---|---|
| *CHR vs. HC* | | | | | | |
| R insula | 48 | 30, -8, 10 | 1.414 | .0006 | 749 | R insula (369)<br>R putamen (142)<br>R superior temporal gyrus (76)<br>R rolandic operculum (33) |
| R supplementary motor area | 6 | 6, -18, 60 | 1.352 | .0011 | 197 | R supplementary motor area (140) |
| L middle frontal gyrus | 9 | -24, 46, 32 | 1.380 | .0009 | 194 | L middle frontal gyrus (138)<br>L superior frontal gyrus, dorsolateral (49) |
| L parahippocampal gyrus | 36 | -28, -8, -30 | 1.295 | .0002 | 112 | L parahippocampal gyrus (42)<br>L fusiform gyrus (24) |
| L superior frontal gyrus | 8 | 6, 32, 50 | -1.821 | .0002 | 1294 | L superior frontal gyrus, medial (396)<br>L supplementary motor area (335)<br>R superior frontal gyrus, medial (258)<br>R supplementary motor area (180)<br>R median cingulate cortex (43)<br>L median cingulate cortex (22) |
| R inferior parietal lobule | 40 | 42, -44, 48 | -1.962 | <.0001 | 1079 | R inferior parietal lobule (631)<br>R superior parietal lobule (205)<br>R supramarginal gyrus (149) |

| | | | | | | R angular gyrus (60) |
| --- | --- | --- | --- | --- | --- | --- |
| | | | | | | R postcentral gyrus (31) |
| L superior parietal lobule | 7 | -22, -66, 52 | -1.658 | .0005 | 577 | L superior parietal lobule (230) |
| | | | | | | L middle occipital gyrus (133) |
| | | | | | | L inferior parietal lobule (128) |
| | | | | | | L precuneus (35) |
| | | | | | | L superior occipital gyrus (33) |
| R middle frontal gyrus | 45 | 44, 32, 32 | -1.806 | .0002 | 208 | R middle frontal gyrus (151) |
| | | | | | | R inferior frontal gyrus, triangular part (57) |
| R precuneus | 5 | 6, -52, 50 | -1.424 | .0021 | 73 | R precuneus (64) |
| *FEP vs. HC* | | | | | | |
| R middle temporal gyrus | 37 | 42, -66, 14 | 1.606 | .0003 | 160 | R middle temporal gyrus (101) |
| R cerebellum | 19 | 26, -72, -20 | 1.302 | .0028 | 83 | R cerebellum, hemispheric lobule VI (76) |
| L precuneus | 7 | -14, -66, 56 | -1.815 | .0002 | 340 | L precuneus (212) |
| | | | | | | L superior parietal lobule (117) |
| R postcentral gyrus | 3 | 38, -30, 46 | -1.752 | .0003 | 289 | R postcentral gyrus (168) |
| | | | | | | R inferior parietal lobule (66) |
| | | | | | | R supramarginal gyrus (36) |
| R middle frontal gyrus | 45 | 50, 34, 32 | -1.895 | .0001 | 222 | R inferior frontal gyrus, triangular part (122) |
| | | | | | | R middle frontal gyrus (100) |

| Long-term SCZ vs. HC | | | | | | |
|---|---|---|---|---|---|---|
| R orbitofrontal cortex | 11 | 2, 40, -22 | 1.824 | .0001 | 1048 | R superior frontal gyrus, medial orbital (282) <br> L gyrus rectus (207) <br> L superior frontal gyrus, medial orbital (200) <br> R gyrus rectus (173) <br> L anterior cingulate cortex (66) <br> R anterior cingulate cortex (11) |
| L middle frontal gyrus | 9 | -30, 28, 46 | 1.663 | .0004 | 224 | L middle frontal gyrus (218) |
| R insula | 48 | 32, 14, 2 | -2.632 | <.0001 | 1234 | R putamen (720) <br> R insula (189) |
| R cerebellum | 30 | 8, -44, -20 | -2.253 | <.0001 | 530 | R cerebellum, hemispheric lobule IV/V (174) <br> R cerebellum, hemispheric lobule III (105) |
| R thalamus | / | 4, -18, 2 | -2.339 | <.0001 | 431 | R thalamus (129) <br> L thalamus (49) |

*Note:* Suprathreshold clusters were identified at p < .005 and cluster size > 50 voxels for the separate analysis for different illness stages. Peak coordinates were reported in the standard template of Montreal Neurological Institute (MNI), and the number of cluster breakdowns (> 10 voxels) was calculated by adding subclusters reported by SDM software. Abbreviations: L = left; R = right; CHR = clinical high-risk state for psychosis; FEP = first-episode psychosis; SCZ = schizophrenia; HC = healthy controls.

Comparative Meta-analysis Across Illness Stages of SCZ**Figure Legends**

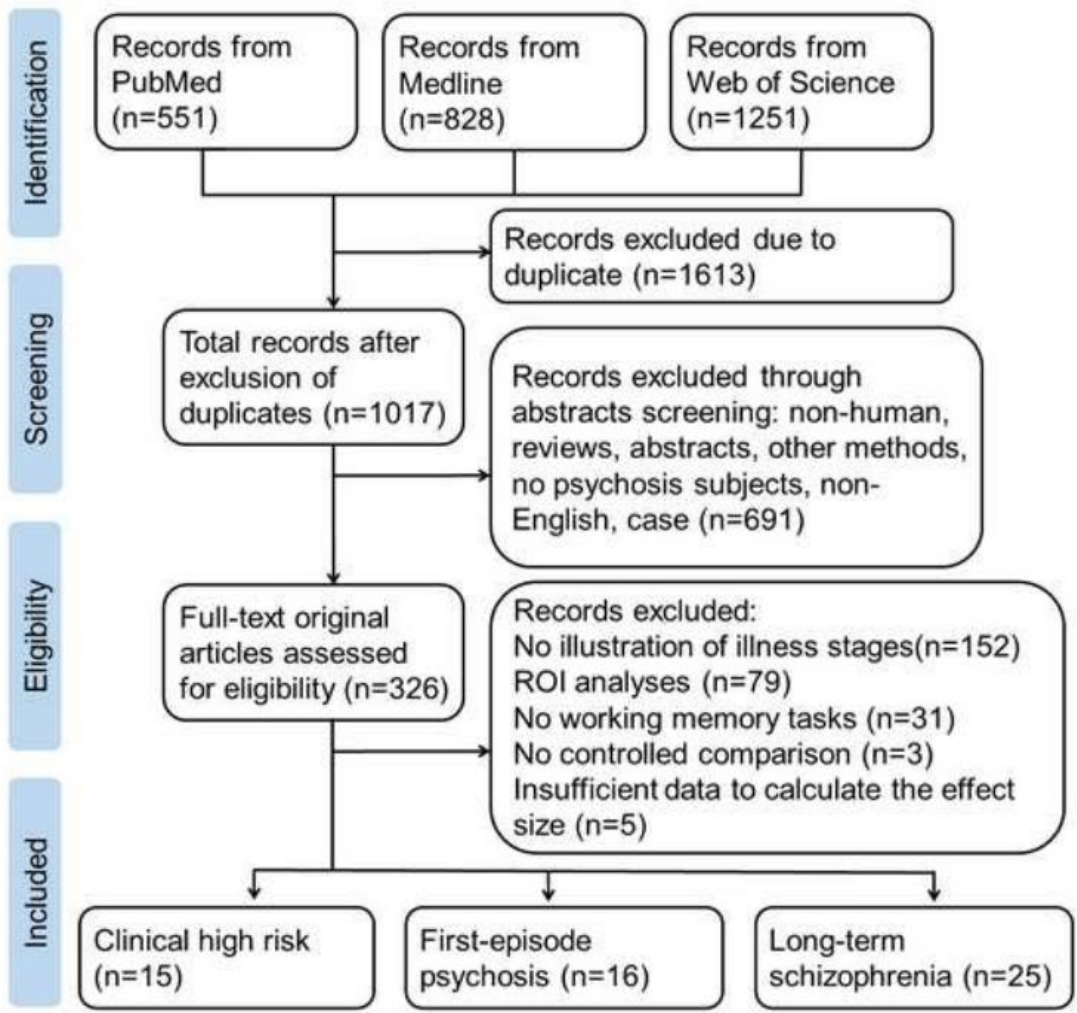

**Figure 1.** Flowcharts of the literature search and selection criteria. *Abbreviations:* ROI = region of interest.

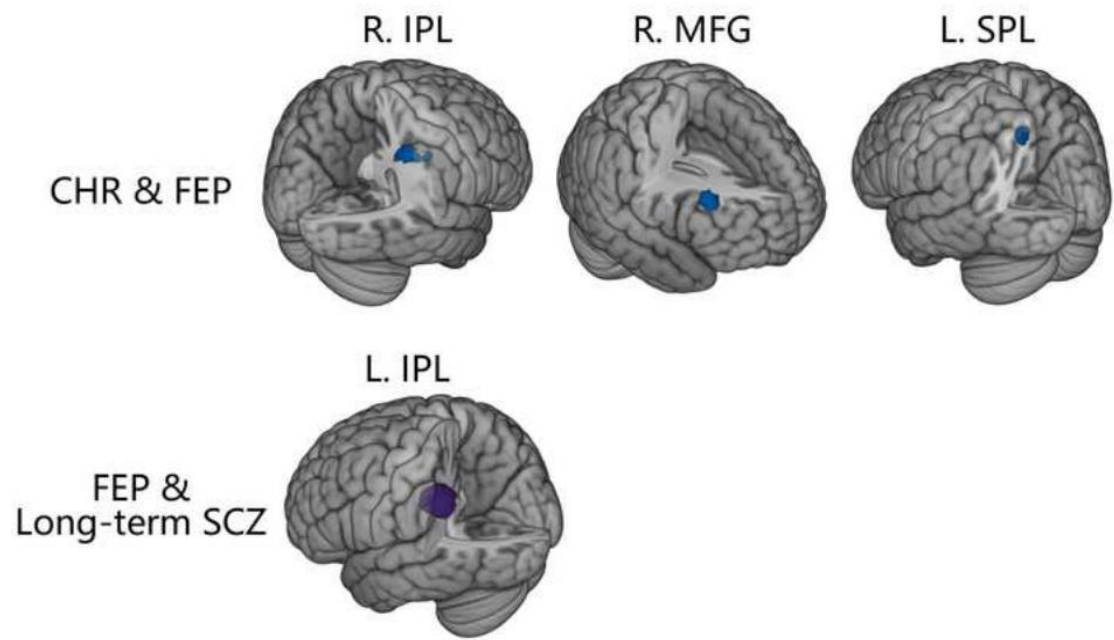

**Figure 2.** Overlapping and comparative findings of different illness stages. Blue, the same brain region that showed decreased activation. Purple, relatively less activation in long-term schizophrenia. *Abbreviations:* L = left; R = right; CHR = clinical high-risk state for psychosis; FEP = first-episode psychosis; SCZ = schizophrenia; IPL = inferior parietal lobule; MFG = middle frontal gyrus; SPL = superior parietal lobule.

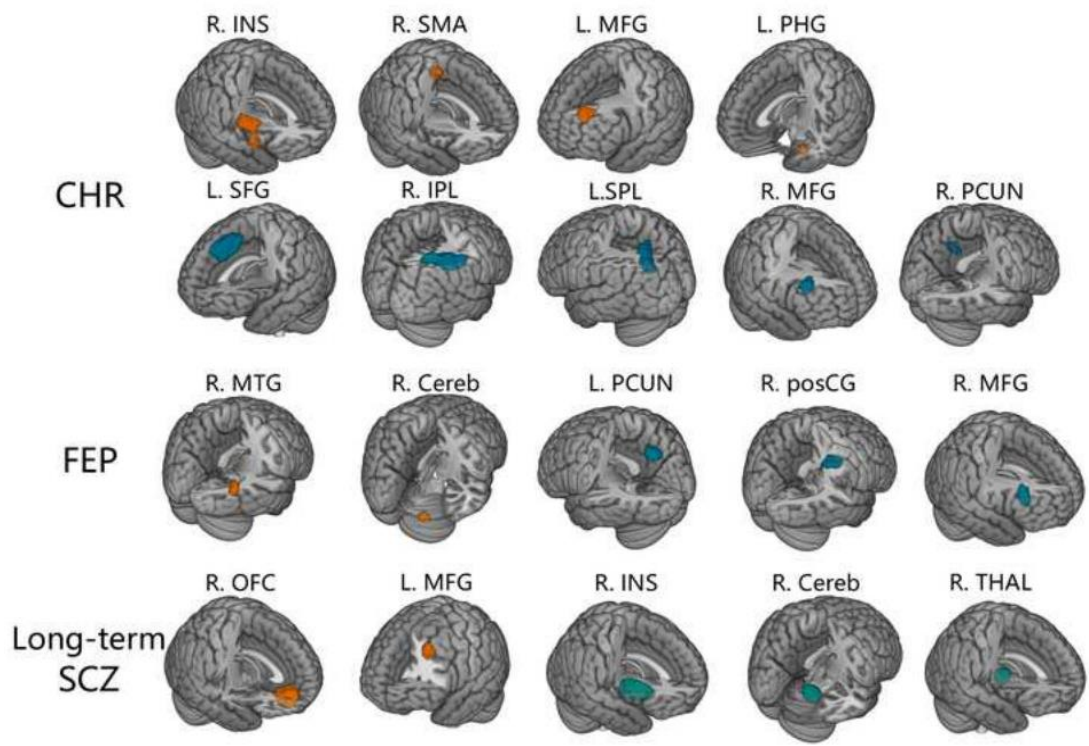

**Figure 3.** Separate findings of different illness stages compared with healthy controls. Orange, brain regions with increased activation. Blue, brain regions with decreased activation. *Abbreviations:* L = left; R = right; CHR = clinical high-risk state for psychosis; FEP = first-episode psychosis; SCZ = schizophrenia; INS = insula; SMA = supplementary motor area; MFG = middle frontal gyrus; PHG = parahippocampal gyrus; SFG = superior frontal gyrus; IPL = inferior parietal lobule; SPL = superior parietal lobule; PCUN = precuneus; MTG = middle temporal gyrus; Cereb = cerebellum; posCG = postcentral gyrus; OFC = orbitofrontal cortex; THAL = thalamus.